\documentstyle[prl,twocolumn,aps,epsf]{revtex}
\begin{document}

\title{Multi-boson effects and the normalization of the two-pion
  correlation function} 

\author{ Q.H. Zhang, P. Scotto and U. Heinz} 

\address{Institut f\"ur Theoretische Physik, Universit\"at 
Regensburg, D-93040 Regensburg, Germany}

\date{\today}

\maketitle

\begin{abstract}

The two-pion correlation function can be defined as a ratio of either
the measured momentum distributions or the normalized momentum space
probabilities. We show that the first alternative avoids certain
ambiguities since then the normalization of the two-pion correlator
contains important information on the multiplicity distribution of the
event ensemble which is lost in the second alternative. We illustrate
this explicitly for specific classes of event ensembles. 
\end{abstract}

PACS numbers: 25.75.-q, 25.75.Gz, 25.70.Pq. 

\section{Introduction}
\label{sec1}

Two-particle Bose-Einstein (BE) interferometry (also known as Hanbury 
Brown-Twiss (HBT) intensity interferometry) as a method for obtaining 
information on the space-time geometry and dynamics of high energy 
collisions has recently received intensive theoretical and
experimental attention. Detailed investigations revealed that
high-quality two-particle correlation data constrain not only the
geometric size of the particle-emitting source but also its dynamical
state at particle freeze-out \cite{Zajc86,Lorstad89,BGJ,APW,Pratt95,He96}.   

Two different definitions of two-pion correlation function are employed 
in the literature
\cite{Zajc86,Lorstad89,BGJ,APW,Pratt95,He96,GKW79,Zajc84,Mark,UA1,Padula,Alex93,CDL,MV97}.
The first starts from the measured invariant $i$-pion inclusive distribution
 \begin{equation}
 \label{1}
  N_i({\bbox{p}_1, \bbox{p}_2,\dots, \bbox{p}_i})
  = E_{\bbox{p}_1}\cdots E_{\bbox{p}_i}
  {1\over \sigma} \frac{d^{3i}\sigma}{d^3p_1 d^3p_2 \cdots d^3p_i} \, ,
 \end{equation}
which is normalized via
 \begin{eqnarray}
 \label{2}
   && \int {d^3p_1\over E_{\bbox{p}_1}}\cdots {d^3p_i\over E_{\bbox{p}_i}}
   N_i(\bbox{p}_1,\dots,\bbox{p}_i) 
 \nonumber\\
   &&\qquad\qquad\qquad = \langle n(n-1)\cdots(n-i+1) \rangle
 \end{eqnarray}
to the $i$th order factorial moment of the pion multiplicity
distribution, and defines the two-particle correlator as
 \begin{equation}
 \label{I}
   C^I({\bbox{p}_1,\bbox{p}_2}) = 
   \frac{N_2(\bbox{p}_1,\bbox{p}_2)}
        {N_1({\bbox{p}_1})N_1({\bbox{p}_2})}\, .
\end{equation}
The second definition instead employs the normalized $i$-pion
production probability
 \begin{equation}
   P_{i}({\bbox{p}_1,\dots,\bbox{p}_i})
   =\frac{N_i({\bbox{p}_1,\dots,\bbox{p}_i})}
         {\langle n(n-1)\cdots(n-i+1)\rangle}
\end{equation}
and defines
 \begin{equation}
 \label{II}
   C^{II}(\bbox{p}_1,\bbox{p}_2) = 
   \frac{P_2({\bbox{p}_1},{\bbox{p}_2})}
        {P_1({\bbox{p}_1})P_1({\bbox{p}_2})}\, .
 \end{equation}
It follows that 
 \begin{equation}
 \label{18}
   C^{II}(\bbox{p}_1,\bbox{p}_2) = 
   \frac{\langle n \rangle^2}{\langle n(n-1)\rangle}\,
   C^{I}(\bbox{p}_1,\bbox{p}_2)\,.
 \end{equation}

Recently, Mi\'skowiec and Voloshin \cite{MV97} argued that the first
definition is preferable because it is based directly on measured
quantities and it is consistent with the often used theoretical
expression
 \begin{equation}
 \label{corr}
  C(\bbox{q,K}) = 1 + 
  {\left|\int d^4x\, S(x,\bbox{K}) \, e^{iq\cdot x} \right|^2
   \over \int d^4x\, S(x,\bbox{p}_1)\, \int d^4y\, S(y,\bbox{p}_2)}\, .
 \end{equation}
Here $\bbox{K}=(\bbox{p}_1+\bbox{p}_2)/2$, 
$\bbox{q}=\bbox{p}_1-\bbox{p}_2$,
$q^0=E_{\bbox{p}_1}-E_{\bbox{p}_2}$, and $S(x,\bbox{K})$ is the  
emission function of the source. In this paper we will stress that the
first definition has the additional advantage that it provides
information not only about {\it the shape of the correlator}, but also
through its normalization about {\it the pion multiplicity 
distribution} which is lost in the second definition. We will show
that it can be exploited to search for multi-pion symmetrization 
effects and may thus be a useful ingredient in the HBT analysis of 
2-pion correlation functions.

\section{A simple example}
\label{sec2}

To illustrate the importance of the normalization of the 2-pion 
correlator let us start with a simple example in which we
consider the following multi-pion states:
 \begin{eqnarray}
 \label{7}
   |\phi\rangle_m &=& A_m \exp [ (\hat B^\dagger)^{m} ] |0\rangle \, ,
 \label{xx1}\\
   \hat B^\dagger &=& \int d^3p\, j(\bbox{p})\, a^\dagger(\bbox{p}).
 \end{eqnarray}
The states (\ref{7}) are normalizable for $m\leq 2$; the norma\-li\-zation
$A_m$ ensures $_m\langle\phi|\phi\rangle_m=1$. For $m=1$, 
$|\phi\rangle_1$ is a standard coherent state \cite{GKW79} with 
$A_1= \exp(-n_0/2)$ and $n_0= \int d^3p \, |j(\bbox{p})|^2$. Then the 
pion multiplicity distribution is of Poisson form,
 \begin{equation}
 \label{9}
   P(n)= \frac{n_0^n}{n!}\exp(-n_0), 
 \end{equation} 
and the two-pion correlators (\ref{I}) and (\ref{II}) are given by
 \begin{equation}
   C^{I}({\bbox{p}_1,\bbox{p}_2})=
   C^{II}({\bbox{p}_1,\bbox{p}_2})= 1\, .
 \end{equation}
For $m=2$ we have $A_2 = (1-4 n_0^2)^{\frac{1}{4}}$, and the pion 
multiplicity distribution is given by
 \begin{equation}
   P(n) = \left\{
 \begin{array}{ccl}
   0 &,&\ n=2k+1~{\rm odd},\\
   (1-4n_0^2)^{\frac{1}{2}}\frac{(2k)!}{(k!)^2}n_0^{2k} &,&\ n=2k~{\rm even}.
 \end{array} \right.
 \end{equation} 
Calculating the correlation functions we find
 \begin{equation}
   C^{I}({\bbox{p}_1,\bbox{p}_2}) = 2+\frac{1}{4n_0^2}, 
   \quad
   C^{II}({\bbox{p}_1,\bbox{p}_2})=1.
\end{equation}
One observes that now $C^{I}$ is different from $C^{II}$ due to the
fact that the pion multiplicity distribution is no longer of Poisson 
form. However, although $|\phi\rangle_2$ is clearly not a coherent
state, $C^{II}$ is again equal to $1$. The use of the second
definition (\ref{II}) thus does not allow to distinguish between 
the states $|\phi_1\rangle$ and $|\phi_2 \rangle$, whereas the first 
definition (\ref{I}) clearly does. One may, of course, argue that the 
only difference between $C^{I}$ and $C^{II}$ is the normalization
factor which can be obtained independently by measuring the pion 
multiplicity distribution. Our point is that important information on
the pion multiplicity distribution can also be extracted directly from
the properly normalized correlation function, and that this opportunity
should not be given away by working with probabilities rather than
directly with the measured cross sections.

\section{Multi-boson effects on the correlator and its normalization}
\label{sec3}

We will now consider a more physical model and show again that the use
of the second definition leads to a loss of interesting information
about the source. It is well known that in relativistic heavy-ion
collisions the pion multiplicity is so large that it may be necessary
to take multi-pion BE correlations into account 
\cite{APW,Pratt95,Zajc,Pratt93,CGZ,ZC1,CZ1,Zhang,Zhang2,SBA97,Pol1,Pol2,Urs}.
In the following we will use a specific class of density matrices for
multi-pion systems \cite{Pratt93,CGZ,ZC1} to study multi-pion BE
correlation effects on the two-pion correlation function, thereby
generalizing the conclusions of Ref.\cite{MV97}. For this class of
ensembles it was shown in \cite{Pratt93,CGZ,ZC1} that, after
including multi-pion correlation effects, the two-pion and single-pion
inclusive distributions can, in the notation of
\cite{Pratt93,CGZ,ZC1}, be written in the following simple form \cite{ZZZ}:  
 \begin{eqnarray}
 \label{12}
   N_1(\bbox{p}) &=& E_{\bbox{p}} H(\bbox{p,p})\, ,
 \\
 \label{13}
   N_2({\bbox{p}_1,\bbox{p}_2})
   &=& E_{\bbox{p}_1} E_{\bbox{p}_2} \bigl[ H(\bbox{p}_1,\bbox{p}_1)
   H(\bbox{p}_2,\bbox{p}_1) 
 \nonumber\\
   &&\qquad\quad + H(\bbox{p}_1,\bbox{p}_2) H(\bbox{p}_2,\bbox{p}_1)
   \bigr]\, ,
 \\
 \label{14}
   H({\bbox{p}_1,\bbox{p}_2}) &=&
   \sum_{i=1}^{\infty} G_i({\bbox{p}_1,\bbox{p}_2}) \, . 
\end{eqnarray}
The $G_i(\bbox{p,q})$ are defined as
 \begin{equation}
 \label{15}
   G_i(\bbox{p},\bbox{q}) = \int \rho(\bbox{p},\bbox{p}_1)
   d^3p_1 \rho(\bbox{p}_1,\bbox{p}_2) \cdots d^3p_{i-1}
   \rho(\bbox{p}_{i-1},\bbox{q}) ,
 \end{equation}
where $\rho(\bbox{p}_i,\bbox{p}_j)$ is a Fourier transform of the source
emission function $g(x,\bbox{K})$:
 \begin{equation}
 \label{16}
   \rho(\bbox{p}_i,\bbox{p}_j) = \int d^4x\, g\left(x,\bbox{K}_{ij}\right)
   \, e^{iq_{ij}\cdot x}\, .
 \end{equation}
Here $\bbox{K}_{ij}{=}(\bbox{p}_i+\bbox{p}_j)/2$ and
$\bbox{q}_{ij}{=}\bbox{p}_i-\bbox{p}_j$, 
$q^0_{ij}{=}E_{\bbox{p}_i}-E_{\bbox{p}_j}$. Inserting the expressions
(\ref{12},\ref{13}) into Eq.~(\ref{I}) one obtains 
 \begin{equation}
 \label{17}
   C^{I}(\bbox{p}_1,\bbox{p}_2) = 1 +
   \frac{H(\bbox{p}_1,\bbox{p}_2) H(\bbox{p}_2,\bbox{p}_1)}
        {H(\bbox{p}_1,\bbox{p}_1) H(\bbox{p}_2,\bbox{p}_2)}\,.
 \end{equation}
This correlator goes to 1 as $\bbox{q}\to\infty$ and to 2 as
$\bbox{q}\to 0$. (Final state interactions are neglected here.) Thus
even dramatic multi-boson effects as discussed below do not
affect the intercept of the correlator $C^I$ --- although they change
the multiplicity distribution towards Bose-Einstein form they do not
lead to genuine phase coherence. 

Explicit expressions for the pion multiplicity distribution and its
first two moments $\langle n\rangle$, $\langle n(n-1)\rangle$ for the
model studied here can be found in \cite{Pratt93,CGZ,ZC1,CZ1}. Since
$H(\bbox{p}_1,\bbox{p}_2)=H^*(\bbox{p}_2,\bbox{p}_1)$, the second   
term in Eq.~(\ref{17}) is always positive, ensuring that $C-1$ is
positive definite. The normalization conditions (\ref{2}) therefore
imply that for the class of systems studied here and in
\cite{Pratt93,CGZ,ZC1,CZ1} one has always $\langle n(n-1)\rangle > 
\langle n\rangle^2$ (see Fig.~3 below). Obviously, Eq.~(\ref{17}) can
therefore not apply to systems with multiplicity distributions $P(n)$
which give $\langle n(n-1)\rangle \leq \langle n\rangle^2$
(e.g. for systems with fixed event multiplicity
\cite{Pratt93,Zhang,Urs}). 

The structure of (\ref{17}) permits to introduce, in analogy to
(\ref{16}), a modified source distribution $S(x,\bbox{K})$ via 
 \begin{equation}
 \label{20}
   H(\bbox{p}_1,\bbox{p}_2) = \int d^4x\, S(x, \bbox{K})\, 
   e^{iq\cdot x}
 \end{equation}
such that the correlator (\ref{17}) can be written in the form
(\ref{corr}). $S(x,\bbox{K})$ is related to the original source
distribution $g(x,\bbox{K})$ via Eqs.~(\ref{14})-(\ref{16}). It
includes all higher order multiparticle BE symmetrization
effects. When interpreting measured single particle spectra and 
two-particle correlations one must keep in mind that the extracted
information on the source {\it corresponds to the effective source
  distribution $S(x,\bbox{p})$ rather than to the emission function
  $g(x,\bbox{p})$}. The following example shows that these two
functions can differ considerably; but we will also see that an
important clue as to how much they differ will be provided by the
normalization of the correlator.

As shown in \cite{Pratt93,CGZ,ZC1,CZ1} the recursion relations 
for the functions $G_i$ in (\ref{15}) can be solved analytically
for the class of model ensembles studied here if the following 
source distribution $g(x,\bbox{p})$ is assumed:
 \begin{eqnarray}
 \label{21}
   g(\bbox{r},t,\bbox{p}) 
   &=& n_0 \left( {1\over 2\pi R^2} \right)^{3/2}
       \exp\left(-\frac{\bbox{r}^2}{2R^2}\right)
 \nonumber\\
   && \times \left( \frac{1}{2\pi \Delta^2} \right)^{3/2}
      \exp\left(-\frac{\bbox{p}^2}{2 \Delta^2}\right)\, 
      \delta(t)\, .
 \end{eqnarray} 
$g(\bbox{r},t,\bbox{p})$ is the {\em Wigner} density of the source in
the absence of multi-pion symmetrization effects. It is obtained in 
\cite{ZC1,CZ1} by folding the Wigner densities of individual Gaussian
wavepackets with a classical phase-space distribution $\rho_{\rm
  class}$ for their centers (Eq.~(19) in \cite{CH94}; see also
\cite{ZWSH}). The parameters $R,\Delta$ in (\ref{21}) are thus
combinations of the wavepacket width $\sigma$ with the spatial and
momentum space widths $R_{\rm class}$ and $\Delta_{\rm class}$ of
$\rho_{\rm class}$ (see Eqs.~(20,21) in \cite{Weal}). While the width
parameters $R_{\rm class}, \Delta_{\rm class}$ of the classical
distribution $\rho_{\rm class}$ are unconstrained, the widths
$R,\Delta$ of the Wigner density $g(\bbox{r},t,\bbox{p})$ which result
from the folding procedure always satisfy the quantum mechanical
uncertainty relation $R\Delta \geq \hbar/2$.

The {\em input} multiplicity distribution is taken to be Poissonian as
in (\ref{9}); its mean value $n_0$ can be interpreted as the mean pion
multiplicity in the absence of Bose-Einstein correlations
\cite{Pratt93,CGZ}. By inspection of Eqs.~(\ref{14})-(\ref{16}) one
easily convinces oneself that the instantaneous character of
(\ref{21}) carries over to the effective source distribution
$S(x,\bbox{p})$. Using the analytical expressions from
Refs.~\cite{Pratt93,CGZ,ZC1,CZ1} we compute     
 \begin{equation}
   N_1(\bbox{p}) = E_{\bbox{p}}\,H(\bbox{p},\bbox{p})
   = E_{\bbox{p}} \int d^4x\, S(x,\bbox{p}) 
 \end{equation} 
%
\begin{figure}[h]\epsfxsize=8cm
\centerline{\epsfbox{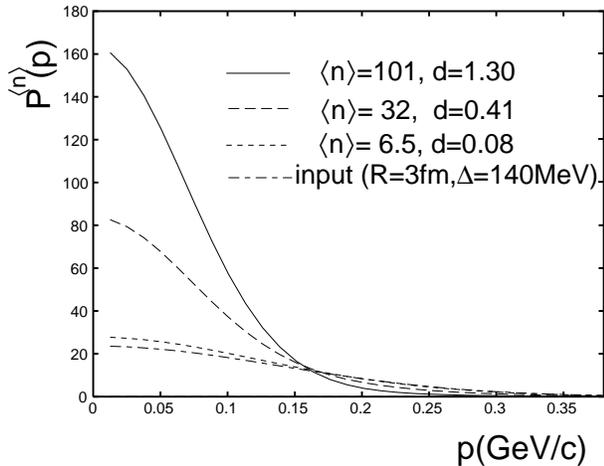}}
\caption{ 
   Multi-pion correlation effects on the single-pion spectrum. The
   dash-dotted line corresponds to the input distribution $\int d^4x\,
   g(x,\bbox{p})$ with $R=3$ fm and $\Delta=140$ MeV.
   The other lines correspond to the measured distribution for various
   values of the average pion multiplicity $\langle n\rangle$ per
   event. Also given are the corresponding average phase space
   densities $d$ (see text). } 
\end{figure}
%
\noindent
as well as the normalized single-pion probability distribution in
momentum space
 \begin{equation}
   P_1^{\langle n \rangle}(\bbox{p}) =
   \frac{E_{\bbox{p}}}{\langle n \rangle} \int d^4x\, S(x,\bbox{p}) =
   \frac{E_{\bbox{p}}}{\langle n \rangle} \sum_{i=1}^{\infty} 
   G_i(\bbox{p,p})\, .
\end{equation}
The mean pion multiplicity $\langle n \rangle$ is given by 
 \begin{equation}
   \langle n \rangle = \int d^3p\, d^4x\, S(x,p) = \sum_{i=1}^{\infty} 
   \int d^3p\, G_i(\bbox{p,p})\, .
 \end{equation}
For the model (\ref{21}) $P_1^{\langle n \rangle}(p)$ is a function of
$p=|\bbox{p}|$ only. It is shown in Fig.~1 for different {\em observed}
average pion multiplicities $\langle n \rangle$. Next to the value
$\langle n \rangle$ we also give the average pion phase-space density
of the system,
 \begin{equation}
 \label{d}
   d = {\langle n \rangle \over (2 R\Delta)^3} \, .
 \end{equation}
One sees that as $d$ increases the pions concentrate in momentum space
at low momenta. This reflects their bosonic nature: pions like to be
in the same state.      

The instantaneous nature of the (effective) emission functions
$g(x,\bbox{p})$ and $S(x,\bbox{p})$ (see (\ref{21})) allows for
inversion of the Fourier transform (\ref{20}): writing $S(x,\bbox{p})=
S(\bbox{r},\bbox{p})\,\delta(t)$ we have
 \begin{equation}
 \label{25}
   S(\bbox{r,K}) = \int {d^3q \over (2\pi)^3}\,  
   H\left(\bbox{K}+{\textstyle{\bbox{q} \over 2}},
          \bbox{K}-{\textstyle{\bbox{q} \over 2}}\right)
   e^{i\bbox{q}\cdot \bbox{r}}\,.
 \end{equation}

We define the normalized source distribution in coordinate space 
$P^{\langle n\rangle}(\bbox{r})$ via 
 \begin{eqnarray}
 \label{26}
   && P^{\langle n \rangle}(\bbox{r})
    = {\int d^3K\, S(\bbox{r,K}) \over
       \int d^3K\, d^3r\, S(\bbox{r,K})}
 \nonumber\\
   && \qquad
    = {1\over \langle n \rangle} \int {d^3K\, d^3q\, \over (2\pi)^3}\, 
       H\left(\bbox{K}+{\textstyle{\bbox{q} \over 2}},
              \bbox{K}-{\textstyle{\bbox{q} \over 2}}\right)\,
        e^{i\bbox{q}\cdot \bbox{r}} \, .
 \end{eqnarray}
%
\begin{figure}[h]\epsfxsize=8cm
\centerline{\epsfbox{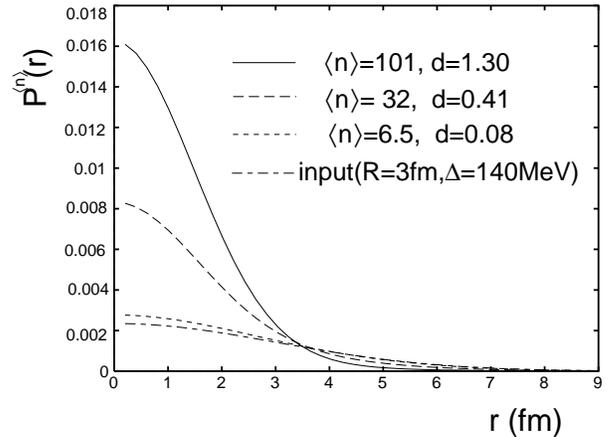}}
\caption{ 
  Multi-pion correlation effects on the spatial pion distribution. 
  The dash-dotted line corresponds to the input momentum distribution
  $\int dt\,d^3p\,g(x,\bbox{p})$ with parameters $R=3$ fm and 
  $\Delta=140$ MeV. The other lines correspond to the effective
  spatial distribution which would be extracted from HBT measurements,
  for various values of the average pion multiplicity $\langle
  n\rangle$ per event. Also given are the corresponding average phase
  space densities $d$ (see text). }
\end{figure}
%
\noindent
Due to the spherical symmetry of (\ref{21}) it is a function of
$r=|\bbox{r}|$ only. The function $H$ in (\ref{25},\ref{26}) is
known analytically \cite{Pratt93,CGZ,ZC1,CZ1} to be a simple Gaussian
in $\bbox{q}$, rendering the Fourier transform trivial. The resulting
$P^{\langle n \rangle}(r)$ is shown in Fig.~2 for different 
pion phase-space densities. One sees that with increasing phase-space
density the multi-pion BE correlations also lead to a concentration of
the pions in coordinate space. The fact that multi-pion BE
correlations lead to a reduction of the HBT radius has been observed
previously \cite{Zajc,CGZ,Urs}. The radius extracted from HBT
interferometry reflects the typical length scale of $P^{\langle n
  \rangle}(r)$; it is always smaller than the input geometric radius
$R$ of the source and depends on the mean pion multiplicity per event.

\section{The high density limit}
\label{sec4}

Taking the limit of a highly condensed Bose gas, $d = {\langle n
  \rangle \over (2 R\Delta)^3} \rightarrow \infty$ \cite{Comm1}, the
multiplicity distribution and 1- and 2-particle spectra can be
determined analytically \cite{ZC1}:  
 \begin{eqnarray}
 \label{27}
   P(n) &=& \frac{\langle n \rangle^n}{(\langle n \rangle +1)^{n+1}}\, ,
 \\
 \label{28}
   N_1(\bbox{p}) &=& E_{\bbox{p}} 
   {\langle n\rangle \over (2\pi\Delta_{\rm eff}^2)^{3/2}}\,
   \exp\left(-\frac{\bbox{p}^2}{2\Delta_{\rm eff}^2}\right)\,, 
 \\
 \label{29}
   N_2(\bbox{p}_1,\bbox{p}_2) &=& 2\, N_1(\bbox{p}_1)\,
   N_1(\bbox{p}_2)\, ,
 \\
 \label{30}
  \Delta_{\rm eff}^2 &=& {\Delta \over 2R} \leq \Delta^2\, .
 \end{eqnarray}
In this limit the correlation functions are 
 \begin{equation}
  C^{I}(\bbox{p}_1,\bbox{p}_2) = 2, \qquad
  C^{II}(\bbox{p}_1,\bbox{p}_2) = 1.
 \end{equation} 
Multi-pion BE correlations change the original Poisson distribution
into the Bose-Einstein multipli\-city distribution
(\ref{27}). Correspondingly, ${\langle n(n-1) \rangle \over \langle
  n\rangle^2}$ changes from 1 to 2. In (\ref{18}) this change exactly
compensates the fact that the correlator $C^I$ no longer decays as a
function of $\bbox{q}{=}\bbox{p}_1{-}\bbox{p}_2$, and from the
resulting $C^{II}\equiv 1$ one might thus be misled to conclude
(incorrectly) that the source exhibits phase coherence.   

In Fig.~3 we show the ratio ${\langle n\rangle^2 \over \langle
n(n-1)\rangle}$ as a function of the average pion phase-space
density. In the lower diagram we plot it as a function of the ratio
of input parameters $n_0/(2R\Delta)^3$, in the upper diagram we use as
a measure of the phase-space density the analogous ratio formed with
the {\em measured} average multiplicity $\langle n \rangle$. For low
phase space densities and large systems ($2R\Delta \gg 1$) ${\langle
  n\rangle^2 \over \langle n(n-1)\rangle}$ is close to 1; BE
symmetrization effects on the observed multiplicity distribution are
then negligible, and it becomes equal to the Poissonian input
distribution with $\langle n(n-1)\rangle = \langle n\rangle^2 =
n_0^2$. For smaller systems ($2R\Delta \simeq 1$) the observed
multiplicity distribution becomes non-Poissonian, with ${\langle
  n\rangle^2 \over \langle n(n-1)\rangle} < 1$, even in the limit of
vanishing multiplicity, $d = \langle n \rangle /(2R\Delta)^3 \to 0$:
for the model discussed here one finds analytically   
 \begin{equation}
 \label{lim}
   \lim_{n_0\to 0} {\langle n(n-1) \rangle \over \langle n\rangle^2}
   = 1 + {1\over (2R\Delta)^3} \leq 2\, .
 \end{equation}

\begin{figure}[h]\epsfxsize=8cm
\centerline{\epsfbox{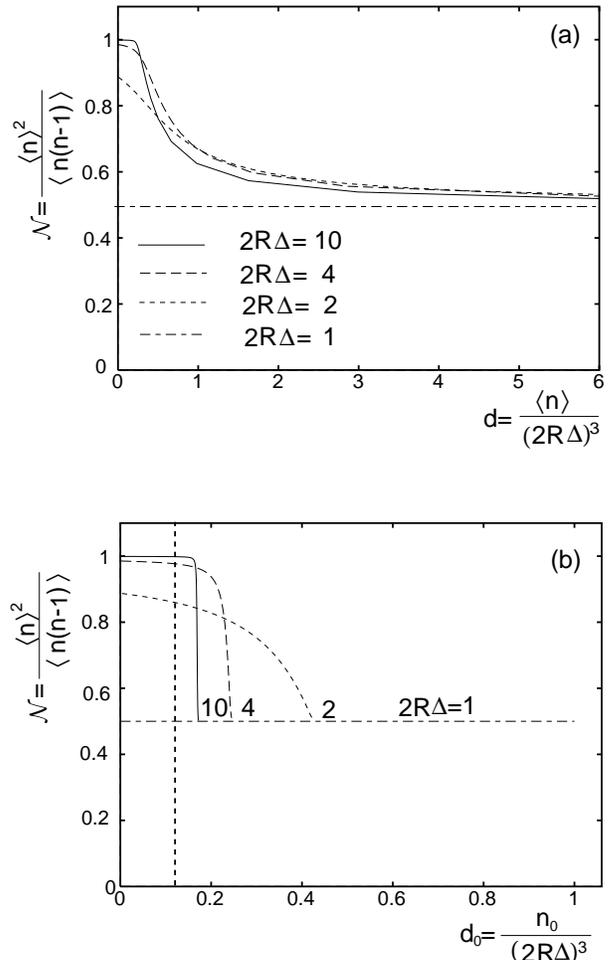}}
\caption{ 
  Multi-pion correlation effects on the normalization factor as a
  function of the average pion phase-space density, for different
  total phase space volumes $2R\Delta=1,\, 2,\, 4,\, 10$ as
  indicated. The corresponding input parameters were $R= 0.7,\, 1.4,\,
  3,$ and 7 fm, respectively, as well as $\Delta=140$ MeV in all
  cases. For details see text.}
\end{figure}

For large phase-space densities $d{>}1$, the ratio ${\langle
  n\rangle^2 \over \langle n(n-1)\rangle}$ decreases, eventually 
approaching for $d\to \infty$ the value 0.5 which reflects 
Bose-Einstein statistics. The critical phase-space density for the 
transition from Poisson statistics with $\langle n(n-1)\rangle =
\langle n\rangle^2$ to Bose-Einstein statistics with $\langle
n(n-1)\rangle = 2 \langle n\rangle^2$ depends on the total phase-space
volume $(2R\Delta)^3$, but for $2R\Delta \gg 1$ it occurs near 
$d\simeq 0.3$. For $d\gg 1$, multi-boson symmetrization effects
become dominant. The Bose condensation limit is reached at a finite
critical value for the mean input multiplicity $n_0$: 
 \begin{equation}
 \label{nc}
   n_0 \to n_c = \left( R\Delta + {\textstyle{1\over 2}}\right)^3 
   \geq 1\, .
 \end{equation}
As $n_0$ approaches the critical value (which for large systems
$2R\Delta \gg 1$ corresponds to $n_c/(2R\Delta)^3 \approx {1\over 8}$), 
the observed mean multiplicity $\langle n\rangle$ and phase-space
density $d$ (as well as the total energy) go to infinity
\cite{Pratt93}. In this sense the limit $n_0\to n_c$ here is analogous
to the limit $\mu_\pi \to m_\pi$ in a thermalized pion gas of infinite
volume in the grand canonical formalism.

\section{Normalized correlation functions from experiment}
\label{sec5}

A direct experimental determination of the phase-space density $d=
\langle n \rangle/(2R\Delta)^3$ in high energy collisions is not
easy. The sources created in such collisions feature strong collective
expansion \cite{Pratt95,He96}, and therefore only small fractions of
the total collision region (so-called regions of homogeneity)
contribute effectively to the two-particle correlator \cite{He96}. 
This means that in the parametrization (\ref{21}) we should use 
$R^2 = R_{\rm hom}^2 + 1/(4\Delta^2)$ where $R_{\rm hom}$ is the (pair
momentum dependent) homogeneity radius which, in the absence of strong
multi-pion effects, is equal to the HBT radius parameter $R_{\rm
  HBT}$. Without more detailed model studies it is then, however,
unclear what fraction of the total observed multiplicity $\langle n
\rangle$ comes from a single such homogeneity region.  

On the other hand Fig.~3 suggests that, within our model class
of event ensembles, the ratio ${\langle n\rangle^2 \over \langle
  n(n-1)\rangle}$ is a useful indicator for the average phase-space
density $d$ in the source and thereby also for the expected multi-pion
symmetrization effects on the 1- and 2-particle spectra which need to
be taken into account in an extraction of the source size from HBT
measurements. 

In the experiment one usually fits the two-particle correlator with
the functional form 
 \begin{equation}
   C^{\rm exp}(\bbox{p}_1,\bbox{p}_2) = 
   C^{\rm exp}(\bbox{q,K}) = 
   {\cal N} \left( 1{+}\lambda |f(\bbox{q,K})|^2\right) ,
 \end{equation}
where the function $f$ vanishes as $\bbox{q}\to \infty$. Obviously,
${\cal N}$ depends on the chosen definition of the correlation
function: For the definition (\ref{I}) the normalization is
always ${\cal N}^{I}=1$ (see Eq.~(\ref{17})), while for the definition
(\ref{II}) it is ${\cal N}^{II} = {\langle n\rangle^2 \over \langle
  n(n-1)\rangle}$ (see Eq.~(\ref{18})). But in both cases ${\cal N}$
is well-defined and thus should not be treated as a free fit
parameter. Therefore we now discuss shortly an algorithm for the 
experimental construction of the two-particle correlator which is
guaranteed \cite{Comm2} to give the correct value for ${\cal N}$,
{\em without relying on an actual measurement of the multiplicity
distribution}.  

We write the single-pion inclusive distribution as
 \begin{equation}
   N_1(\bbox{p}) = \frac{E_{\bbox{p}}}{N_{\rm ev}} 
                   \sum_{i=1}^{N_{\rm ev}} \nu_i(\bbox{p})\, .
 \end{equation}
$N_{\rm ev}$ is the total number of collision events, and
$\nu_i(\bbox{p})$ is the number of pions with momentum $\bbox{p}$  
in collision $i$. The two-particle distribution can be expressed as
 \begin{equation}
 \label{35}
   N_2(\bbox{p}_1,\bbox{p}_2) = 
   {E_{\bbox{p}_1}\, E_{\bbox{p}_2} \over N_{\rm ev}} 
   \sum_{i=1}^{N_{\rm ev}} \tilde\nu_{i,i}(\bbox{p}_1,\bbox{p}_2)\, ,
\end{equation}
where $\tilde\nu_{i,i}(\bbox{p}_1,\bbox{p}_2)$ is the number of pion
pairs with momenta $(\bbox{p}_1,\bbox{p}_2)$ in collision event $i$,
and the double index $i$ indicates that both particles are from the
same event. These definitions satisfy the normalization conditions
(\ref{2}). While $N_2(\bbox{p}_1,\bbox{p}_2)$ is constructed by
selecting pion pairs from the same events, the denominator
$N_1(\bbox{p}_1)N_1(\bbox{p}_2)$ can be generated by combining pion
pairs from different events
\cite{Kop74,Zajc84,Mark,UA1,Alex93,CDL,MV97}. The proper prescription
is 
 \begin{equation}
 \label{36}
   N_1(\bbox{p}_1) N_1(\bbox{p}_2) = 
   {E_{\bbox{p}_1} \, E_{\bbox{p}_2} \over N_{\rm ev}(N_{\rm ev}-1)}
   \sum_{i,j=1 \atop i\ne j}^{N_{\rm ev}}
   \tilde\nu_{i,j}(\bbox{p}_1, \bbox{p}_2)
 \end{equation}
where $\tilde\nu_{i,j}(\bbox{p}_1, \bbox{p}_2) = \nu_i(\bbox{p}_1)
\nu_j(\bbox{p}_2)$. One easily checks that 
 \begin{equation}
   \int \frac{d^3p_1}{E_{\bbox{p}_1}} \frac{d^3p_2}{E_{\bbox{p}_2}}
        N_1(\bbox{p}_1) N_1(\bbox{p}_2) = \langle n \rangle^2\, .
 \end{equation}
The ratio of (\ref{35}) and (\ref{36}) thus gives the properly
normalized correlator $C^{I}$. The above equations are true for
unbiased events. Trigger biases and limited experimental acceptances
can induce residual correlations in the event-mixed ``background''
(\ref{36}) which must be corrected for separately (see \cite{Zajc84}
for an extensive discussion).

For large $N_{\rm ev}$ the evaluation of (\ref{36}) is very time
consuming; it also leads to a statistically unnecessarily accurate
result for the denominator in (\ref{I}). In practice one can
live with fewer event pairs for event mixing, by replacing in
(\ref{36}) the number $N_{\rm ev}$ by a much smaller number $N'_{\rm
  ev}$. As long as $N'_{\rm ev}(N'_{\rm ev}-1) \gg N_{\rm ev}$ one can
still ensure that the contribution of the denominator to the
statistical error of the final correlation function is negligible
\cite{Comm2}. 
 
The correlator $C^{II}$ differs from $C^{I}$ only by the different
normalization. It can be constructed by taking the ratio of the
following two expressions \cite{Zajc84}:
 \begin{eqnarray}
 \label{38}
   P_2(\bbox{p}_1,\bbox{p}_2) &=& 
   {E_{\bbox{p}_1} \, E_{\bbox{p}_2} \over N^c_{\rm pairs}}
   \sum_{i=1}^{N_{\rm ev}} \tilde\nu_{i,i}(\bbox{p}_1,\bbox{p}_2) \, ,
 \\
 \label{39}
   P_1(\bbox{p}_1) P_1(\bbox{p}_2) &=& 
   {E_{\bbox{p}_1} \, E_{\bbox{p}_2} \over N^u_{\rm pairs}}
   \sum_{i,j=1 \atop i\ne j}^{N_{\rm ev}}
   \tilde\nu_{i,j}(\bbox{p}_1, \bbox{p}_2) \,.
 \end{eqnarray}
$N^c_{\rm pairs}$ and $N^u_{\rm pairs}$ are the total numbers 
of ``correlated'' and ``uncorrelated'' pion pairs, respectively:
 \begin{eqnarray}
   N^c_{\rm pairs} &=& N_{\rm ev} \cdot \langle n(n-1)\rangle \,, 
 \\
   N^u_{\rm pairs} &=& N_{\rm ev}(N_{\rm ev}-1) \cdot 
                       \langle n \rangle^2 \,.
 \end{eqnarray}

\section{Conclusions}
\label{sec6}

We have shown that in principle the normalization of the two-particle
Bose-Einstein correlation function contains valuable information on
the the multiplicity distribution of the event ensemble. Both
theoretically and experimentally the absolute normalization of the
correlation function should thus be controlled as well as possible. We
presented a variant of a previously suggested experimental algorithm
\cite{MV97} for the construction of the correlator which guarantees
correctly normalized correlators. Within a specific model class of
event ensembles which recently received extensive theoretical
attention we showed that in systems with large pion phase-space
densities multi-pion symmetrization effects can lead to interesting
measurable effects on the normalization of the correlator. We suggest
a careful study of this normalization as an alternate method for
searching for strong multi-pion symmetrization effects in
high-multiplicity hadronic and heavy-ion collisions.

\acknowledgements

The authors thank D. Mi\'skowiec, U. Wiedemann, C. Slotta, and
T. Cs\"org\H o  for helpful discussions. Q.H.Z. gratefully
acknowledges support by the Alexander von Humboldt Foundation. The
work of U.H. and P.S. was supported in part by DFG, BMBF, and
GSI. U.H. would like to thank the Institute for Nuclear Theory in
Seattle for its hospitality and for providing a stimulating
environment while this work was completed. 
 


\begin{thebibliography}{100}
\bibitem{Comm1}
  Since $2R\Delta \geq 1$, the limit indicated in the text can
  only be achieved by letting $\langle n \rangle \to \infty$ 
  or $n_0 \to n_c$ (see Eq.~(\protect\ref{nc})).
\bibitem{Comm3}
  Combining this result with Eqs.~(\ref{17}) and (\ref{18}) gives a
  correlator of the form $C^{II}(\bbox{q,K}) = {\cal N} (1 + \vert
  f(\bbox{q,K})\vert^2)$ with ${\cal N} = 1/[1 + (2R\Delta)^{-3}]
  \approx 1 - {1\over (2R\Delta)^3}$. A similar result was also
  obtained in \cite{Urs} in the limit of small phase-space densities,
  although with quite different assumptions about the event ensemble.
  In particular, in \cite{Urs} the events were assumed to have a fixed
  pion multiplicity $n$. For this reason the factor ${\cal N}$ in
  \cite{Urs} cannot be associated with the pion multiplicity 
  distribution $P(n)$; still, the leading dependence of the
  normalization of the correlator on multi-pion effects enters through
  the same factor involving the phase-space volume $(2R\Delta)^3$ of
  the source.
\end{thebibliography}
\end{document}